\documentclass[]{elsart3p}

\usepackage{amsmath,amssymb}
\journal{Physics Letters B}
\usepackage{graphicx}
\begin{document}

\begin{frontmatter}
\title{Scalar field cosmological models with finite scale factor singularities}

\author{Francesco Cannata}
\address{INFN, Via Irnerio 46,40126 Bologna,
Italy}\ead{Francesco.Cannata@bo.infn.it}
\author{Alexander Y. Kamenshchik\corauthref{cor}}\corauth[cor]{Corresponding author}
\address{Dipartimento di Fisica and INFN, Via Irnerio 46,40126 Bologna,
Italy\\
L.D. Landau Institute for Theoretical Physics of the Russian
Academy of Sciences, Kosygin str. 2, 119334 Moscow, Russia}\ead{Alexander.Kamenshchik@bo.infn.it}
\author{Daniele Regoli}
\address{Dipartimento di Fisica, Via Irnerio 46,40126 Bologna,
Italy}\ead{astapovo@gmail.com}
\begin{abstract}  
We construct a scalar field based cosmological model, possessing a cosmological singularity characterized 
by a finite value of the cosmological radius and an infinite scalar curvature. Using the methods of the qualitative theory of differential equations, 
we give a complete description of the cosmological evolutions in the model under consideration.  
There are four classes of evolutions, two of which have finite lifetimes, while the other two undergo an infinite expansion.  
\end{abstract}
\begin{keyword}
scalar fields in cosmology\sep cosmological singularities\sep future of the universe 
\PACS 98.80.Cq\sep 98.80.Jk
\end{keyword}

\end{frontmatter}
\section{Introduction}
The discovery of the phenomenon of cosmic acceleration \cite{cosmic} 
has stimulated a study of a huge variety of cosmological models, based 
on perfect fluids, scalar fields, tachyons etc \cite{dark}. 
The cosmological models based on scalar  fields  have a long history, being used 
for exploration of possible inflationary scenarios \cite{inflation} and for description of dark energy 
\cite{quintessence}. This has attracted a special attention to the technique of reconstruction of scalar 
potentials reproducing a given cosmological evolution \cite{reconstruct}.

On the other hand this development of model-designing art has revealed  cosmological evolutions
possessing various types of singularities, sometimes very different from the traditional Big Bang and 
Big Crunch. The most popular between them is, perhaps, the Big Rip cosmological singularity 
\cite{Star-rip,Rip} arising in superaccelerating models driven by some kind of phantom matter \cite{phantom}.
Other types of singularities are sudden singularities \cite{sudden}, Big Brake \cite{we-tach}, and so on \cite{other}.
Here we would like to study the singularities which are close to the known Big Bang, Big Crunch and Big Rip singularities,
but arising at finite values of the cosmological factor (different from zero and infinity as well). 
Similar singularities were recently considered in \cite{freeze}.

In the present paper we construct potentials which can drive the cosmological evolution towards (or from) such 
singularities. Combining qualitative and numerical methods we study the set of possible cosmological histories 
in the suggested models to show that the presence of such singularities in a cosmological model under consideration 
depends essentially on initial conditions and that the same model can accomodate qualitatively different 
cosmological scenarios.  

The structure of the paper is the following: in Sec. 2 we construct some potentials corresponding to evolutions 
with ``soft'' singularities. In the third section we analyze their dynamics. The conclusion is devoted to an interpretation 
of the obtained results. 

\section{Construction of scalar field potentials}

We shall consider flat Friedmann models with the metric 
\begin{equation}
ds^2 = dt^2- a^2(t)dl^2.
\label{Fried}
\end{equation}
The Hubble parameter $h(t) \equiv \dot a/a$ 
satisfies the Friedmann equation
\begin{equation}
h^2 = \varepsilon,
\label{Fried1}
\end{equation}
where $\varepsilon$ is the energy density and a convenient normalization of the Newton constant is chosen.
Differentiating equation (\ref{Fried1}) and using the energy conservation equation
\begin{equation}
\dot{\varepsilon} = -3h(\varepsilon + p),
\label{en-conserv}
\end{equation}
where $p$ is the pressure, one comes to
\begin{equation}
\dot{h} = -\frac32 (\varepsilon + p).
\label{h-dot}
\end{equation}
If the matter is represented by a spatially homogeneous minimally coupled scalar field, then
the energy density and the pressure are given by the formulae
\begin{equation}
\varepsilon = \frac12\dot{\phi}^2 + V(\phi),
\label{energy}
\end{equation}
\begin{equation}
p = \frac12\dot{\phi}^2 - V(\phi),
\label{pressure}
\end{equation}
where $V(\phi)$ is a scalar field potential.
Combining equations (\ref{Fried1}), (\ref{h-dot}), (\ref{energy}), (\ref{pressure})
we have
\begin{equation}
V = \frac{\dot{h}}{3} + h^2,
\label{poten}
\end{equation}
and
\begin{equation}
\dot{\phi}^2 = -\frac23 \dot{h}.
\label{phi-dot}
\end{equation}
Equation (\ref{poten}) represents the potential as a function of time $t$. Integrating equation (\ref{phi-dot})
one can find the scalar field as a function of time. Inverting this dependence we can obtain the time parameter as
a function of $\phi$ and substituting the corresponding formula into equation (\ref{poten}) one arrives to the uniquely
reconstructed potential $V(\phi)$. It is necessary to stress that this potential reproduces a given cosmological evolution
only for some special choice of initial conditions on the scalar field and its time derivative.
  
It is known that the power-law cosmological evolution is given by the Hubble parameter $h(t) \sim \frac{1}{t}$. 
We shall look for a ``softer'' version of the cosmological evolution given by the law 
\begin{equation}
h(t) = \frac{S}{t^{\alpha}},
\label{soft}
\end{equation}
where $S$ is a positive constant and $0 < \alpha < 1$. At $t = 0$ a singularity is present, but it is different from the 
traditional Big Bang singularity. Indeed, integrating we obtain
\begin{equation}
\ln \frac{a(t)}{a(0)} = \frac{S}{1-\alpha} t^{1-\alpha}.
\label{soft1}
\end{equation}
If $t > 0$ the right-hand side of Eq. (\ref{soft1}) is finite and hence one cannot have $a(0) = 0$ in the left-hand side of this equation, because it would  imply a contradiction, making $\ln\frac{a(t)}{a(0)}$ divergent. Hence $a(0) > 0$, while 
\begin{equation}
\dot{a} = a(0)\frac{S}{t^{\alpha}}\exp\left(\frac{S}{1-\alpha}t^{1-\alpha}\right) \xrightarrow[t\rightarrow 0]{}\infty .
\label{soft2}
\end{equation}
This type of singularity can be called soft Bing Bang singularity because the cosmological radius is finite (and non-zero) while 
its time derivative, the Hubble variable and the scalar curvature are singular. 
It is interesting to note that when $t \rightarrow \infty$ both $a(t)$ and $\dot{a}(t)$ tend to infinity, but 
they do not encounter any cosmological singularity because the Hubble variable and its derivatives tend to zero.

Let us reconstruct the potential of the scalar field model, producing the cosmological evolution (\ref{soft}) using the 
technique described above. 
Eq. (\ref{phi-dot}) gives 
\begin{equation}
\dot\phi=\pm\sqrt{\frac{2}{3}\alpha S}\;t^{-\frac{\alpha+1}{2}}.
\label{phi-dot1}
\end{equation}
We shall choose the positive sign, without loosing generality.
Integrating, we get
\begin{equation}
\phi(t)=\sqrt{\frac{2}{3}\alpha S}\;\frac{2t^{\frac{1-\alpha}{2}}}{1-\alpha},
\label{phi}
\end{equation}
up to an arbitrary constant.
Inverting the last relation we find 
\begin{equation}
t(\phi)=\left(\left(\frac{3}{2\alpha S}\right)^{1/2}\frac{1-\alpha}{2}\phi\right)^{\frac{2}{1-\alpha}}.
\end{equation}
Hence, using Eq. (\ref{poten}) we obtain
\begin{equation}
V(\phi)=\frac{S^2}{\left(\sqrt{\frac{3}{2\alpha S}}\frac{1-\alpha}{2}\phi\right)^{\frac{4\alpha}{1-\alpha}}}-\frac{\alpha S}{3\left(\sqrt{\frac{3}{2\alpha S}}\frac{1-\alpha}{2}\phi\right)^{\frac{2(\alpha+1)}{1-\alpha}}},
\end{equation}
This potential provides the cosmological evolution (\ref{soft}) if initial conditions compatible with Eqs. (\ref{phi-dot1}) and 
(\ref{phi}) are chosen. Naturally, there are also other cosmological evolutions, generated by other initial conditions, which 
will be studied in the next section.

\section{The dynamics of the cosmological model with $\alpha = \frac12$}

In order to achieve some simplification of calculations we shall consider a particular model, namely the one with the choice 
$\alpha = \frac12$.  In this case 
\begin{equation}
a(t)=a(0)e^{2S\sqrt{t}},
\label{a-new}
\end{equation}
and 
\begin{equation}
V(\phi)=\frac{16S^4}{(\sqrt{3}\phi/2)^4}-\frac{32S^4}{3(\sqrt{3}\phi/2)^6}.
\label{poten-new}
\end{equation}
The Klein-Gordon equation reads
\begin{eqnarray}
\ddot\phi+3\dot\phi\ {\rm sign}(h)\sqrt{\frac{1}{2}\dot\phi^2+\frac{16S^4}
{(\sqrt{3}\phi/2)^4}-\frac{32S^4}{3(\sqrt{3}\phi/2)^6}}\nonumber\\
-\frac{32\sqrt{3}S^4}{(\sqrt{3}\phi/2)^5}+\frac{32\sqrt{3}S^4}{(\sqrt{3}\phi/2)^7}=0.
\label{Klein-Gordon}
\end{eqnarray}
This equation is equivalent to the dynamical system  

\begin{equation}
\left\{
\begin{array}{ll}
\dot\phi=&x,\\
\dot x=&-3x\ {\rm sign}(h)\sqrt{\frac{x^2}{2}+\frac{16S^4}{(\sqrt{3}\phi/2)^4}-\frac{32S^4}{3(\sqrt{3}\phi/2)^6}} \\
&+\frac{32\sqrt{3}S^4}{(\sqrt{3}\phi/2)^5}-\frac{32\sqrt{3}S^4}{(\sqrt{3}\phi/2)^7}.
\label{dyn1}
\end{array}\right.
\end{equation}
The qualitative analysis of dynamical systems in cosmology was presented in detail in \cite{qual-dyn}.

First of all, let us notice that the system has two critical points:
\mbox{$\phi = \pm \frac{2}{\sqrt{3}}, x = 0$.}
We consider the linearized system around the point with the positive value of $\phi$:
\begin{equation}\left\{
\begin{array}{ll}
\dot{\varphi}=& x,\\
\dot x =& -3x\ {\rm sign}(h)\frac{4}{\sqrt{3}}S^2+32\cdot 3 S^4\varphi,
\end{array}\right.
\label{dyn2}
\end{equation}
where $\varphi\equiv \phi-2/\sqrt{3}$.

The Lyapunov indices for this system are (for $h > 0$)
\begin{equation}
\lambda_1=-8\sqrt{3}S^2,
\label{lambda1}
\end{equation}
\begin{equation}
\lambda_2=4\sqrt{3}S^2.
\label{lambda2}
\end{equation}
For negative $h$, corresponding to the cosmological contraction, the signs of $\lambda_1$ and $\lambda_2$ are changed. 
The eigenvalues are real and have opposite signs, hence both the critical points are saddles. The universe being in one of these 
two saddle points means that it undergoes a de~Sitter expansion or contraction, according to the sign of $h$, with the 
value of $h = h_0$ given by 
\begin{equation}
h_0 = \frac{4S^2}{\sqrt{3}}.
\label{deSitter}
\end{equation}
For each saddle point there are four separatrices
which separate four classes of trajectories in the phase plane $x,\phi$ corresponding to four 
types of cosmological evolutions.

\begin{figure}[p]
\centering
\includegraphics[scale=1]{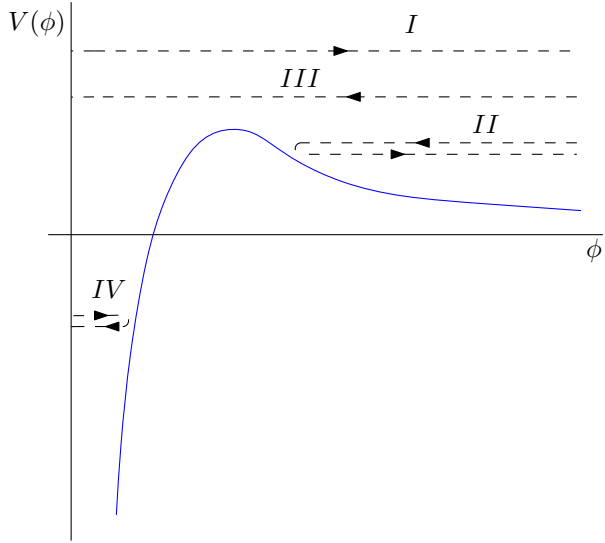}
\caption{Plot of the potential $V(\phi)$ as given by Eq. (\ref{poten-new}). Here only the positive $\phi$-axis is shown, which is enough to understand the behaviour thanks to the parity of the function. We have also delineated the four different dynamical behaviours of the scalar field, clearer to understand taking into account the phase portrait (see Fig.\ref{fig2}).}\label{fig1}
\end{figure}

\begin{figure}[p]
\centering
\includegraphics[scale=1]{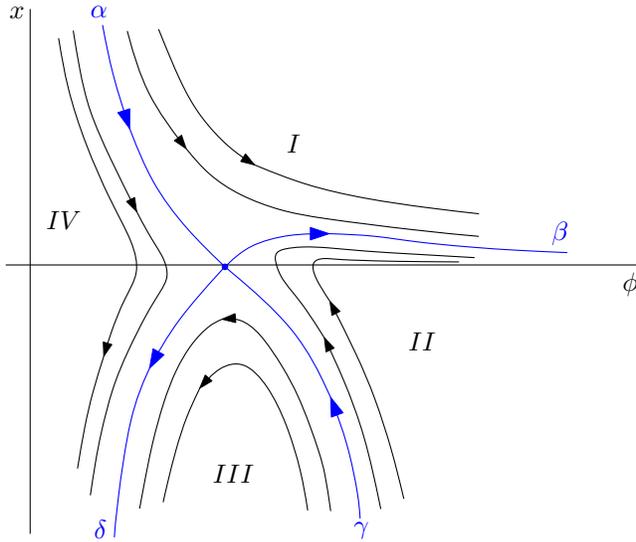}
\caption{Positive $\phi$-axis phase portrait for the dynamical system (\ref{dyn1}) with $h\geq 0$. The four separatrices of the saddle point ($\alpha,\beta,\gamma,\delta$) individuate four regions (\emph{I,II,III,IV}) with different behaviours of the trajectories, as explained in the text.}\label{fig2}
\end{figure}

In order to simplify the study of the dynamics let us note that the potential is an even function of the scalar field $\phi$ and that the saddle points are also symmetrical with respect to the $x$ axis. Thus, it is sufficient to consider only one of this saddle points. We shall carry out our qualitative analysis taking into 
account both Fig.~\ref{fig1}, giving the form of the potential, and  Fig.~\ref{fig2}, representing the phase portrait in the plane $\phi,x$. 

First let us consider trajectories which begin at the moment $t = 0$, when the initial value of the scalar field is infinite, its 
potential is equal to zero and the time derivative of the scalar field is infinite and negative. In terms of the Fig.~\ref{fig1} it means that we consider the motion of the point beginning at the far right on the slope of the potential hill and moving towards the left (i.e. towards the top of the hill) with an infinite initial velocity.  Such a motion for $h > 0$ describes a universe born from the 
standard Big Bang singularity. Further details of this evolution depend on the asymptotic ratio between absolute values of 
$\dot\phi$ and $\phi$ at $t \rightarrow 0$. If this ratio is smaller than some critical value then the scalar field does not 
reach  the top of the hill and at some moment it begins to roll down back to the right. During this process of rolling down the scalar field 
increases, the potential is decreasing and the velocity $\dot{\phi}$ becomes positive and increasing. However the universe expansion works as a friction and at some moment its influence becomes dominant causing an asymptotic damping to zero of the velocity. The universe expands infinitely with $h(t) \rightarrow 0$. In the phase portrait (Fig.~\ref{fig2}) such trajectories 
populate the region $II$. This region is limited by the separatrices $\beta$ and $\gamma$. The first one corresponds to the positive (for $h > 0$) eigenvalue $\lambda_2$, while $\gamma$ corresponds to $\lambda_1$.   
If the ratio $\dot\phi/\phi$ has the critical value, then the scalar field reaches asymptotically the top of the hill of the potential, meaning that the universe becomes asymptotically de~Sitter: in the phase portrait it 
is nothing but the curve $\gamma$.

When the ratio introduced above is larger than the critical one we encounter a different regime. In this case the scalar field 
passes with non-vanishing velocity the top of the hill and begins to roll down in the abyss on the left. The absolute value of the negative velocity $\dot{\phi}$ is growing, 
while the potential becomes negative and at some moment the total energy density of the scalar field vanishes together with the 
Hubble parameter $h$: this means that the universe starts contracting. This contraction provides the growing of the absolute 
value of the velocity of the scalar field and the kinetic term again becomes larger then the potential one. Moreover, both the terms in the Klein-Gordon equation increase the absolute value of $\dot{\phi} < 0$. One can easily show that the regime in which the time derivative $\dot{\phi}$ becomes equal to $-\infty$ at some \emph{finite} value of $\phi$ is impossible, because it implies a contradiction 
between the asymptotic behaviour of different terms in Eq.~(\ref{Klein-Gordon}). Thus, the universe tends to the singularity 
squeezing to the state with the value of $\phi$ equal to zero and an infinite time derivative $\dot{\phi}$. To understand which 
kind of  singularity the universe encounters, we need  some detail about the behaviour of the scalar field. 
Let us suppose that, approaching the singularity at some moment $t_0$, the scalar field behaves as
\begin{equation}
\phi(t) = \phi_0(t_0-t)^{\mu},
\label{appr}
\end{equation}
where $0 < \mu < 1$. Then the first and second time derivatives are 
\begin{equation}
\dot{\phi}(t) = -\mu\phi_0(t_0-t)^{\mu-1},\ \ \ddot{\phi}(t) = \mu(\mu-1)\phi_0(t_0-t)^{\mu-2}.
\label{appr1}
\end{equation}
The potential behaves as 
\begin{equation}
V = -\frac{2048 S^4}{81\phi_0^6(t_0-t)^{6\mu}}.
\label{appr-pot} 
\end{equation}
To have the Hubble variable well defined we require that the kinetic term is larger than the absolute value of the  negative potential term 
(\ref{appr-pot}), i.e. 
$
2\mu-2 \leq -6\mu,
$
or $\mu \leq \frac14$. Now two opposite cases may hold: (i) the friction term in the Klein-Gordon equation could dominate the potential term or (ii) the opposite situation. For (i) to be the right case,
one should require $2\mu -2 < -7\mu$ or $\mu < \frac29$. In this situation the asymptotic behaviour of the second time derivative 
of $\phi$ should be equal to that of the friction term, or, in other words $\mu-2 = 2\mu-2$, that is $\mu = 0$, which 
obviously is not relevant. Thus we have to consider the range $\frac29 < \mu \leq \frac14$. In this case the potential term 
should be equal to the second time derivative of $\phi$, which implies:
\begin{equation}
\mu = \frac14
\label{mu}
\end{equation}
and 
\begin{equation}
\phi_0 = 4\sqrt{\frac{S}{3}}.
\label{phi0}
\end{equation}
Substituting the values of $\mu$ and $\phi_0$ into the expression for $h(t)$,
we obtain 
\begin{equation}
h(t) = -\frac{S}{\sqrt{t_0-t}}.
\label{apprh}
\end{equation}
Thus, we see that the singularity we are approaching is of the soft Big Crunch type.

\begin{figure}[p]
\centering
\includegraphics[scale=1]{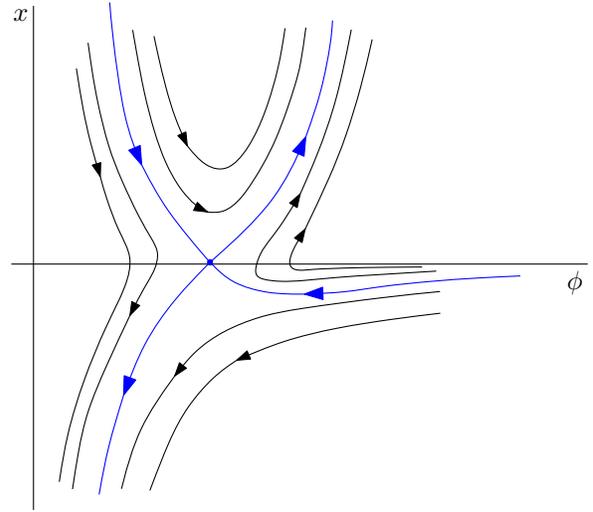}
\caption{Phase portrait for the dynamical system under discussion (Eq. (\ref{dyn1})) with $h\leq 0$, i.e. describing evolutions of the universe characterized by contraction. As for the $h$-positive case, we can see the four separatrices of the saddle point and the corresponding four regions of different dynamical behaviours of the trajectories.}\label{fig3}
\end{figure}

In the phase portrait (Fig.~\ref{fig2}) these trajectories occupy the region $III$ limited by the separatrices $\gamma$ and $\delta$. The cosmological 
evolutions run from the Big Bang singularity to the soft Big Crunch one. However, the Fig.~\ref{fig2} is not sufficient to describe the 
complete behaviour of the universe under consideration, because at some moment the Hubble variable changes sign and we 
should turn to the Fig.~\ref{fig3},  giving the phase portrait for the contracting universe $h < 0$.  Note that increasing the velocity 
$\dot{\phi}$ with which the scalar field overcomes the top of the hill, implies delaying the time when the point 
of maximal expansion of the universe is reached. 

The third regime begins from the soft Big Bang singularity, when the scalar field is equal to zero and its time derivative is 
infinite and positive. In Fig.~\ref{fig1} that situation is represented by the point climbing from the abyss to the top of the hill. If the velocity term 
is not high enough, at some moment the field stops climbing and rolls back down. During this fall the Hubble variable changes
sign and the universe ends its evolution in the soft Big Crunch singularity. The corresponding 
trajectories belong to the region $IV$ of our phase plane, bounded by the separatrices $\delta$ and $\alpha$. The universe 
has its finite  life time between the soft Big Bang and the soft Big Crunch singularities. The situation when the scalar field arrives exactly to the 
top of the hill and stops corresponds to the separatrix $\alpha$. 

The fourth set of cosmological trajectories is generated by the scalar field climbing from the abyss and overcoming 
the top of the hill with the subsequent infinite expansion: the scalar field is infinitely growing and the Hubble parameter 
tends to zero. These trajectories occupy the region $I$ and our original cosmological evolution (\ref{soft}) belongs to this 
family.     

\section{Conclusion}
Let us sum up our results. Wishing to describe a cosmological evolution beginning from (or ending in) a singularity characterized by a nonvanishing
initial (or final) cosmological radius and an infinite value of the scalar curvature due to the infinite value of the Hubble parameter, we have 
constructed a scalar field potential, providing such an evolution. Then, using the methods of qualitative analysis of the 
differential equations, we have shown that the proposed model accomodates four different classes of cosmological evolutions,
depending on initial conditions. Numerical simulations have confirmed our predictions. 

The main results of this work are: (i) the realization 
of a concrete cosmological model with a scalar field, where finite cosmological radius singularities are present and (ii) 
the complete description of all the possible evolutions of this model depending on the initial conditions. It is important to remark that, given 
the fixed scalar field potential, one has different types of evolutions encountering different kinds of singularities.     
The trajectories, belonging to the regions $I$ and $II$ have an infinite time of expansion, while the trajectories belonging 
to the regions $III$ and $IV$ begin and end their histories in the singularities and have a finite lifetime.  

Let us note, that there are some studies \cite{sudden,other,freeze} devoted to the general analysis of various kinds 
of new cosmological singularities. In this work we were not looking for an exhaustive classification of different possible 
cosmological models possessing   singularities, but rather we wanted to study in a complete way a particular cosmological model, having some interesting properties.

\end{document}